\documentclass[aps,prl,showpacs,twocolumn,preprintnumbers,amsmath,amssymb]{revtex4}
\usepackage{graphicx}
\usepackage{color}
\usepackage{hyperref}

\usepackage{booktabs}
\usepackage{array}
\newcolumntype{M}{>{\centering\arraybackslash}m{\dimexpr.25\linewidth-2\tabcolsep}}

\newcommand{\s}{\sigma}
\newcommand{\bk}{{\mathbf k}}

\newcommand{\bR}{\mathbf R}
\newcommand{\bS}{\mathbf S}

\newcommand{\be}{\begin{equation}}
\newcommand{\ee}{\end{equation}}

\begin{document}
\title{Phase diagram and excitations of a Shiba molecule}
\author{N. Y. Yao$^{1}$, C. P. Moca$^{2,3}$, I. Weymann$^{4}$, J. D. Sau$^{5}$, M. D. Lukin$^{1}$, E. A. Demler$^{1}$, G. Zar\'{a}nd$^{2}$}
\affiliation{$^{1}$Physics Department, Harvard University, Cambridge, MA 02138, U.S.A.}
\affiliation{$^{2}$BME-MTA Exotic Quantum Phase Group, Institute of Physics, Budapest University of Technology and Economics, H-1521 Budapest, Hungary}
\affiliation{$^{3}$Department of Physics, University of Oradea, 410087, Oradea, Romania}
\affiliation{$^{4}$Faculty of Physics, Adam Mickiewicz University, 61-614, Pozna\'{n}, Poland}
\affiliation{$^{5}$Joint Quantum Institute and Condensed Matter Theory Center, Department of Physics,
University of Maryland, College Park, Maryland 20742, U.S.A.}

\begin{abstract}
We analyze the phase diagram associated with a pair of magnetic impurities trapped in a superconducting host. The natural interplay between Kondo screening, superconductivity and  exchange interactions leads to a rich array of competing phases, whose transitions are characterized by discontinuous changes of the total spin.
Our analysis is based on a combination of numerical renormalization group techniques as well as semi-classical analytics.
In addition to the expected screened and
unscreened  phases, we observe a new molecular doublet phase where the impurity spins are only partially screened by a single extended quasiparticle. 
Direct signatures of the various Shiba molecule states can be observed via RF spectroscopy. 

\end{abstract}

\maketitle

In an ordinary metal, the celebrated Kondo effect describes the scattering of conduction electrons due to magnetic impurities. Below the so-called Kondo temperature ($T_K$), 
the magnetic moment of
a single impurity becomes screened by the electrons \cite{Hewson}, leading to its dissolution and hence, the formation of a Fermi liquid state \cite{Nozieres}. This simple picture can fail when one considers a finite density of impurities. In particular,  conduction-electrons mediate RKKY exchange interactions, $I$, between the impurities and in the limit,  $I\gtrsim T_K$, such interactions can lead to the emergence of either magnetically ordered or spin glass states \cite{Review,Doniach}.
Much of our understanding of this phase transition owes to detailed studies of the two-impurity Kondo model~\cite{JonesVarma, Affleck}. 

Extending 
the two-impurity calculations 
to the case of a superconducting host represents an interesting and active challenge \cite{Poilblanc94,Balatsky95,Hatte97a,Hatte97b,Salkola97,Flatte00,Pan00,Balatsky,Moca08,Fominov11}.  
On the one hand, 
the interplay of superconductivity and magnetic moments can lead to the emergence of exotic phases and excitations. 
Recent results have suggested the possibility of emergent Majorana edge modes at the ends of a magnetic impurity chain situated on the surface of an s-wave superconductor; 
in this system, topological superconductivity arises from the formation of a spin-helix as a result of the 
underlying RKKY interaction \cite{Ruderman, Kasuya, Yosida}.
On the other hand, the presence of magnetic impurities breaks time-reversal symmetry and gradually leads to the destruction of superconductivity.
This breakdown occurs through the appearance of proliferating mid-gap states (so-called Shiba states),  as first observed by Yu, Shiba and Rusinov \cite{Yu, Shiba, Rusinov}. 
In particular, within a simple classical calculation, they demonstrated that a magnetic impurity can bind an anti-aligned quasiparticle, yielding a sub-gap bound state of energy $\epsilon = \Delta - E_b$, where $\Delta$ represents the superconducting gap and $E_b$ the binding energy \cite{Yazdani97, Ji08}. As the binding energy $E_b$ increases (e.g. as a function of increasing exchange coupling), the bound state energy eventually crosses zero, signifying a parity-changing phase transition.

\begin{figure}[b] 
\centering
\includegraphics[width=0.8\columnwidth]{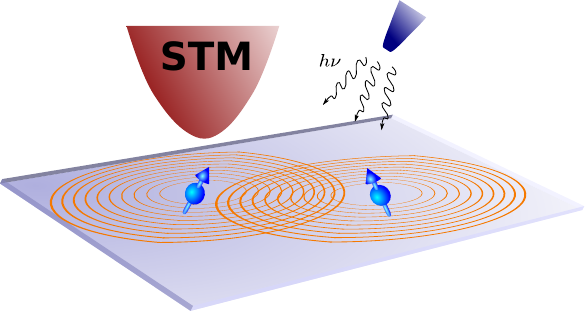}
\caption{
Two magnetic impurities placed on a superconducting surface. RF fields can be used to produce transitions between various molecular states and manipulate them. }
\label{fig:schem}
\end{figure}

With certain modifications, this classical picture  remains qualitatively valid even for 
quantum mechanical spins~\cite{Balatsky, vonDelft.08, Pascal.09, Belzig.13, Bauer,Zitko:2011p609}. Taking into account quantum fluctuations,  the aforementioned parity-changing  transition occurs at a critical point, $(\Delta/T_K)_c$,  when the superconducting gap becomes comparable to the Kondo temperature.
\cite{Hewson}. 
For an  $S=1/2$ impurity,  the spin is essentially free  for $\Delta/T_K  > (\Delta/T_K)_c$  and the associated mid-gap Shiba state remains unoccupied. In this 'free spin' regime, the ground state has spin $S_G = 1/2$. 
In the opposite limit, when $\Delta/T_K <  (\Delta/T_K)_c$, the impurity spin becomes screened by a bound quasiparticle; more specifically, the mid-gap Shiba state becomes occupied and this quasiparticle spin forms a singlet with the impurity spin, leading to an $S_G =0$ ground state. 
This phase transition has recently been observed in mesoscopic circuits, where the strength of the exchange interaction can be tuned by means of a pinch-off gate  electrode \cite{Tarucha}.

In this Letter, through a combination of numerical renormalization group methods and semi-classical analytics, we derive the phase diagram of the two-impurity Kondo model for a superconducting host \footnote{Some results in the limit of $\Delta \ll T_K$ are obtained in \cite{Zitko:2011p609} and agree with the relevant cuts of our phase diagram}. We consider an s-wave superconductor with Hamiltonian,
\begin{eqnarray}
H_{\rm BCS}=  \int \frac{d\bk}{(2\pi)^3}  \Big [ \sum_{\s} \xi_{\bk}c^{\dagger}_{\bk \s}c_{\bk \s}
+\bigl( \Delta c^{\dagger}_{\bk \uparrow}c_{-\bk \downarrow}^{\dagger} +h.c.
\bigr )\Big]
\nonumber
\end{eqnarray} 
coupled, via exchange, to  two identical spin $1/2$ magnetic impurities 
of spin $\bS_1$ and $\bS_2$, 
\be 
H_{\rm int} = {J\over 2}\, \bS_1 \psi^{\dagger}_1{\boldsymbol \s}\,
\psi_1
+
{J\over 2}\, \bS_2 \psi^{\dagger}_2{\boldsymbol \s}\,
\psi_2\;.
\label{eq:H_int0}
\ee
Here, $\psi_1$ and $\psi_2$ are the field operators  at the impurity positions. 
We note that this Hamiltonian captures the essential physics of two experimental systems:  (1)  magnetic 
impurities placed on a superconducting surface (see Fig.~\ref{fig:schem})  \cite{Yazdani97, Ji08,Iavarone:2009p608} and (2) double dot devices attached to superconductors (e.g.~as recently used for Cooper pair splitting) \cite{Csonka,StrunkSplitter}. 
To study the ground state and excitation spectrum of $H_T = H_{\rm BCS} + H_{\rm int}$, we map the problem to a double superconducting chain, and  
analyze it via Wilson's numerical renormalization group (NRG) method \cite{JonesVarma}. Details of our NRG calculation are provided in the Supplementary Material \cite{SuppInfo}. 

We observe that  $H_T$ conserves both parity,  $P$, and  total spin, $S$. In a superconductor, the pairing terms imply that charge is typically only conserved modulo $2$.    However, for  $\Delta=0$ and in the presence of particle-hole symmetry, 
the Wilson chain  possesses a hidden  $SU_c(2)$ charge symmetry \cite{JonesVarma} analogous to
that of
 the Hubbard model \cite{Essler}. 
For a half-filled cubic lattice, 
this charge symmetry is generated by the operators, $Q_x = (Q^+ +Q^-)/2$, 
$Q_y = (Q^+-Q^-)/2i$,
$Q^z =  {1\over 2}\sum_{\s} \int \frac{d\bk}{(2\pi)^3}  ( c^\dagger_{\bk\s}c_{\bk\s}- {1\over 2})$, where $Q^+ =  \int \frac{d\bk}{(2\pi)^3}  c^\dagger_{\bk\uparrow}c^\dagger_{\pi-\bk \hspace{0.5mm}\downarrow }$ and $Q^- =  \big (Q^+\big )^\dagger$
~\footnote{Here $\mathbf \pi$ denotes the corner of the Brillouin zone.}.
Although this symmetry is strictly broken for $\Delta \ne0$,  a hidden  $U_c(1)$ symmetry remains,  leading to a conserved pseudo-charge, $\tilde Q$ \cite{SuppInfo}. Physically, this pseudo-charge can be viewed as the generator of rotations along the superconducting order parameter. 
For the remainder of the text, we will utilize these three quantum numbers ($P$, $S$ and $\tilde Q$)  to classify the eigenstates of the Hamiltonian.

Our NRG calculations reveal the existence of five competing subgap Shiba-molecule states, as depicted in Table~\ref{tab:phases}.
 For  large values of $\Delta$, both of the impurity spins are essentially free. 
 They can form a
singlet state ($S_0$) with spin $S=0$, parity $P=-$, and pseudocharge $\tilde Q =0$, 
or a a triplet state ($T_0$) with $S=1$,  $P=+$, and $\tilde Q =0$.
Similar to the single impurity case, one can also  
create a single (antiferromagnetically) bound quasiparticle. However, in the Shiba molecule case, this quasiparticle is delocalized between the two impurities and can form   either a bonding ($D_+$) or  antibonding state ($D_-$) of spin $S=1/2$, parity $P=\pm$, and pseudo-charge $\tilde Q =1$. 
Finally, it is also possible to induce the binding of two quasiparticles, one to each of the impurities. In this case, one finds a singlet state ($S_2$) with pseudocharge $\tilde Q = 2$. The parity of this state is, rather counterintuitively,  $P = -$, owing to the fermionic nature of the bound quasiparticles.


\begin{table}[t]
 \centering
 \begin{tabular}{MMM}
   \toprule
   \hline
   \multicolumn{2}{c}{State} & $(S,\tilde Q,P)$  \\
   \hline\hline
   \midrule
   \large{$S_0$} & \phantom{n} \includegraphics[width=0.09\textwidth]{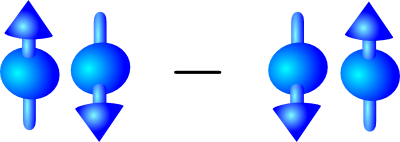} \phantom{n}&  $(0, 0,-)$  \\ 
   \large{$T_0$} & \includegraphics[width=0.03\textwidth]{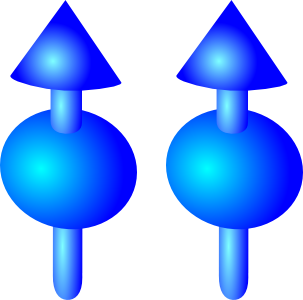} & $(1, 0,+)$ \\
	\large{$D_{\pm}$} & \includegraphics[width=0.12\textwidth]{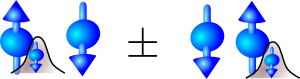} & $({1\over 2},  1,\pm)$ \\
	\large{$S_2$} & \includegraphics[width=0.06\textwidth]{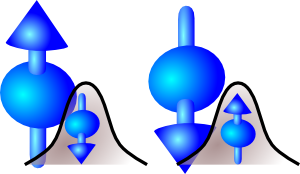} & $(0, 2,-)$ \\
\bottomrule
\hline
 \end{tabular}
 \caption{
\label{tab:phases}
Shiba molecular bound states and their quantum numbers.  Small spins represent quasiparticles bound to the (large) impurity spins. 
}
\end{table}

The competition between these five states leads to a rich Shiba molecule phase diagram. A heuristic understanding of this diagram can be gained by comparing the relative strengths of superconductivity, exchange, and Kondo screening.
 In analogy to the single impurity case, the 
ratio $\Delta/T_K$ characterizes the competition between 
superconductivity and Kondo screening. For $\Delta/T_K\gg1$, Kondo screening is heavily suppressed and the magnetic moments  remain unscreened.
The two impurities do however couple to each other via the Fermi sea of conduction electrons. 
For processes involving quasiparticle excitations close to the Fermi energy, this coupling is characterized  by the overlap $\cal S$ of the  two  waves created at the  impurity locations. For a three dimensional free electron system,  ${\cal S} = \frac{\sin(k_F R)}{k_F R}$, where $R = |\bR_1-\bR_2|$ is the separation between the  impurities and $k_F$ the Fermi momentum. This overlap $\cal S$
is also responsible for the hybridization of the Shiba states at sites 1 and 2, and thus for 
the splitting between the bonding and antibonding states ($D_\pm$).

The impurity spins also interact via  RKKY exchange $I$, which depends on high-energy electron-hole excitations; thus,  the coupling $I$ ought be considered as an independent parameter,  determined by the precise band shape and  the energy dependence of the  exchange coupling, $J$.  
The competition between  RKKY   and  Kondo screening is characterized by the ratio, $I/T_K$ \footnote{In the NRG scheme, a direct interaction between the impurities 
must also be introduced  (see supplementary materials and \onlinecite{JonesVarma}).}.

\begin{figure}[t] 
\centering
\includegraphics[width=0.8\columnwidth]{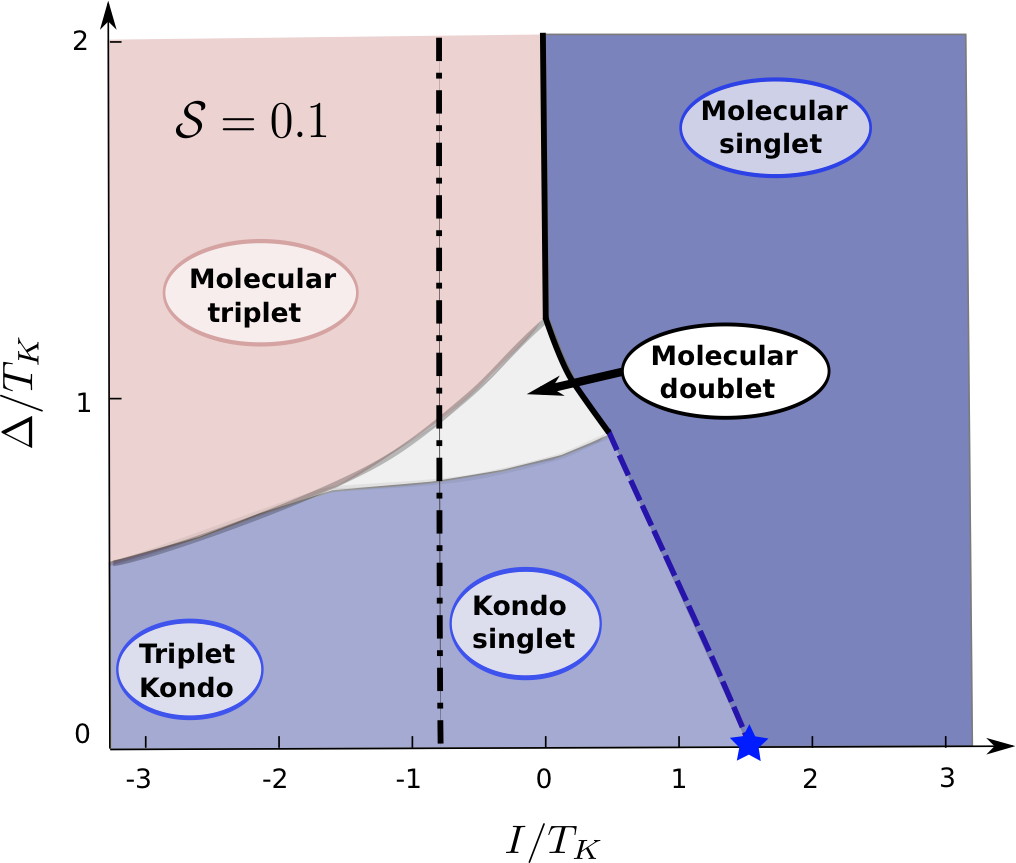}
\caption{
NRG-determined phase diagram for ${\cal S} =0.1$ as function of $I/T_K$
and $\Delta/T_K$. The background colors indicate regions with  $S=1$ (light maroon), 
$S=1/2$ (white) and $S=0$ (blue) ground states. 
The blue dashed line separates the regions with $S_2$ (light blue) and 
$S_0$ (dark blue) ground states.  All phase transitions are first order, except the blue dashed line, which becomes a smooth cross-over in the absence of electron-hole  symmetry. The observed first order transitions are in contrast to the quantum phase transitions observed in the two-channel and two-impurity Kondo models (whose quantum critical point is indicated by the blue star), where local correlation functions exhibit critical behavior with a non-trivial exponent.
 }
\label{fig:phase_S_0}
\end{figure}

The phase diagram obtained via NRG is shown in Fig.~\ref{fig:phase_S_0}. 
We identify four distinct regions, 
each  corresponding one of the states in Table~I: 
(1) For large values of $\Delta/T_K$, the impurities  are  free and the ground state is a 
\emph{molecular triplet} 
($T_0$)
for $I<0$ and   a \emph{molecular singlet} ($S_0$) for $I>0$. As expected, this molecular singlet phase is also observed 
for $I\gg T_K,\Delta$ and extends down to the  $\Delta=0$ axis.   
(2) In the \emph{Kondo singlet} region ($S_2$), $|I|, \Delta \ll T_K$, one recovers strong Kondo correlations, wherein the two impurity spins are basically individually screened 
 by quasiparticles.  For perfect electron-hole symmetry this region is separated from (1) by a first order phase transition (blue dashed line in Fig.~\ref{fig:phase_S_0}), corresponding to both a pseudo-charge jump from $\tilde Q=0$ to $\tilde Q = 2$ as well as a $S_0\to S_2$ singlet-singlet level crossing. When electron-hole symmetry is broken, the transition becomes a smooth cross-over.  (3) Along the  $\Delta=0$ line, the known phase diagram of the two impurity (normal metal) Kondo model is recovered \cite{Doniach}. Here, a quantum critical point (blue star) separates the molecular singlet from the Kondo singlet region.  For any finite $\Delta$, the spectrum is gapped, and this critical point turns into the aforementioned first order transition line.

The nature of the Kondo singlet phase at $\Delta = 0$ gradually changes as one moves toward large, negative exchange interactions.  In particular, for $-I\gg T_K$, the two impurity spins are first bound into a molecular triplet, which is then screened in the even and odd channels  at (typically) two different Kondo temperatures.
This picture survives  for small but finite $\Delta$, although strictly speaking, there is no true Kondo effect for any finite gap; nevertheless, one can still screen the impurity spins for $\Delta\ll T_K$ and a Kondo anomaly is generally observed in the tunneling spectra at intermediate energies,  $\Delta\ll \omega\ll T_K$.

\begin{figure}
\centering
\includegraphics[width=0.8\columnwidth]{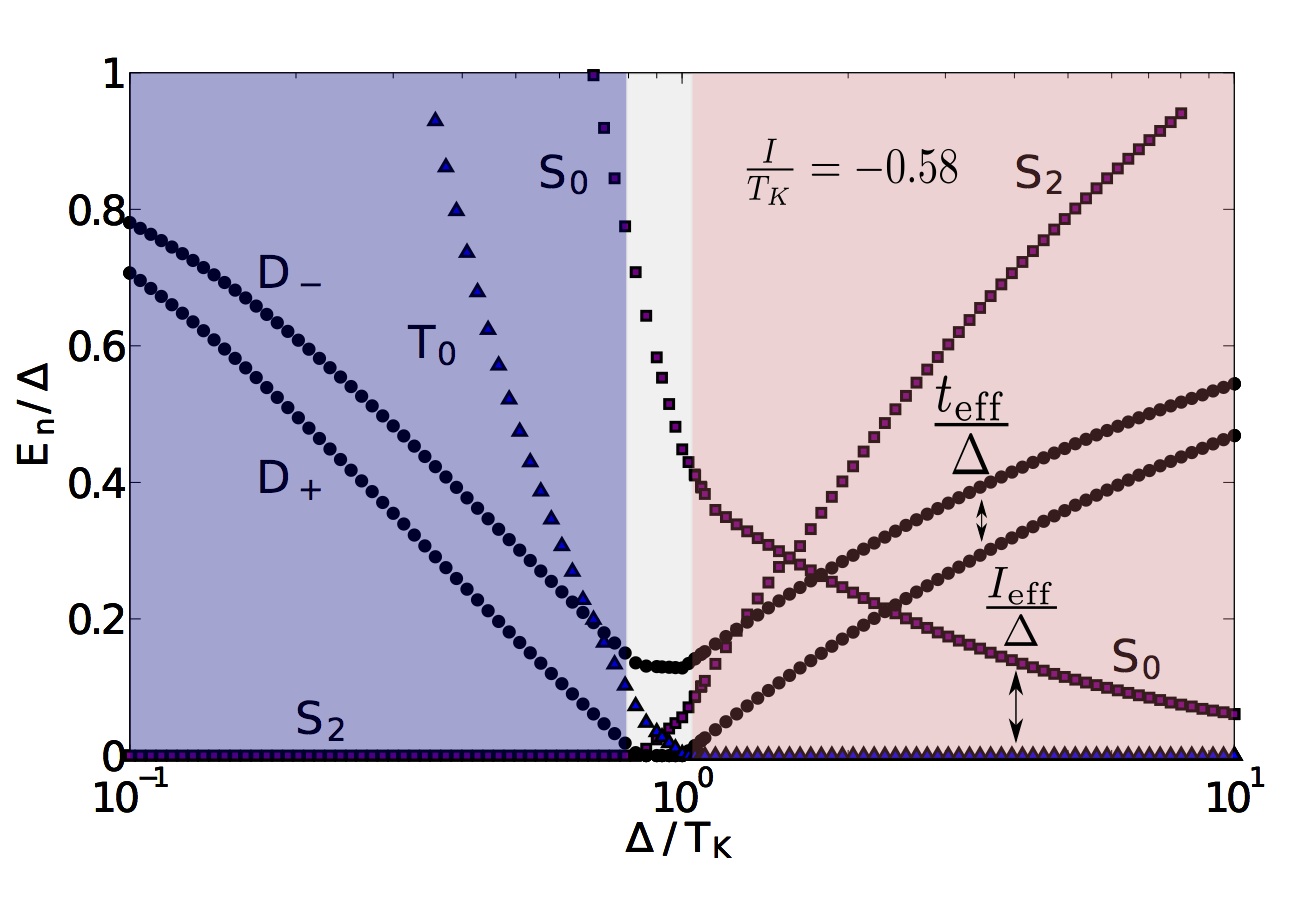}
\caption{
Evolution of the bound states as function of $\Delta/T_K$ for RKKY couplings $I/T_K = -0.58$ and an overlap parameter ${\cal S }= 0.1$ and corresponds 
to the black dashed line in Fig.~2. One observes a phase transition from the 
\emph{individual singlet} state ($S_2$) into the \emph{molecular doublet} phase ($D_+$) and then another 
 transition to the \emph{molecular triplet} phase ($T_0$). 
 The  effective RKKY interaction can be extracted as the splitting 
between the $S_0$ and $T_0$ states: $I_{\rm eff}= E_{{S_0}}-E_{T_0}$. We can 
also estimate the effective hopping $t_{\rm eff}$ as the separation between the two 
molecular doublet  states, $D_\pm$.
 }
\label{fig:cut}
\end{figure}

(4) Finally, and most strikingly, for ${\cal S} \ne 0$ a new $S=1/2$ phase emerges for $\Delta \sim T_K $ and $I\approx 0$. We term this phase the  \emph{molecular doublet} ($D_+$). It can be understood as follows: 
For $\Delta \gg T_K$ each of the two spins  can bind a single excited quasiparticle.  For ${\cal S} = 0$ the energy of these bound states 
are identical; however, for ${\cal S} \ne 0$ these states can hybridize to form molecular bonding and antibonding states $D_\pm$. As one decreases
the ratio $\Delta/T_K$, the energy of the  $D_\pm$ states moves towards zero until $D_+$ first crosses (zero) and becomes the ground state. This transition is  accompanied by a charge-parity flip and 
a spin transition from $S=1\to 1/2$. 
Further decreasing $\Delta/T_K$  lowers the energy of the two-bound-quasiparticle state until  a second charge parity transition to the $S_2$ singlet occurs.  These level crossings and the  evolution of the excitation spectrum along the  vertical  dash-dotted 
line in Fig.~\ref {fig:phase_S_0} is shown in Fig.~\ref{fig:cut}.

\begin{figure}
\center
\includegraphics[width=3.in]{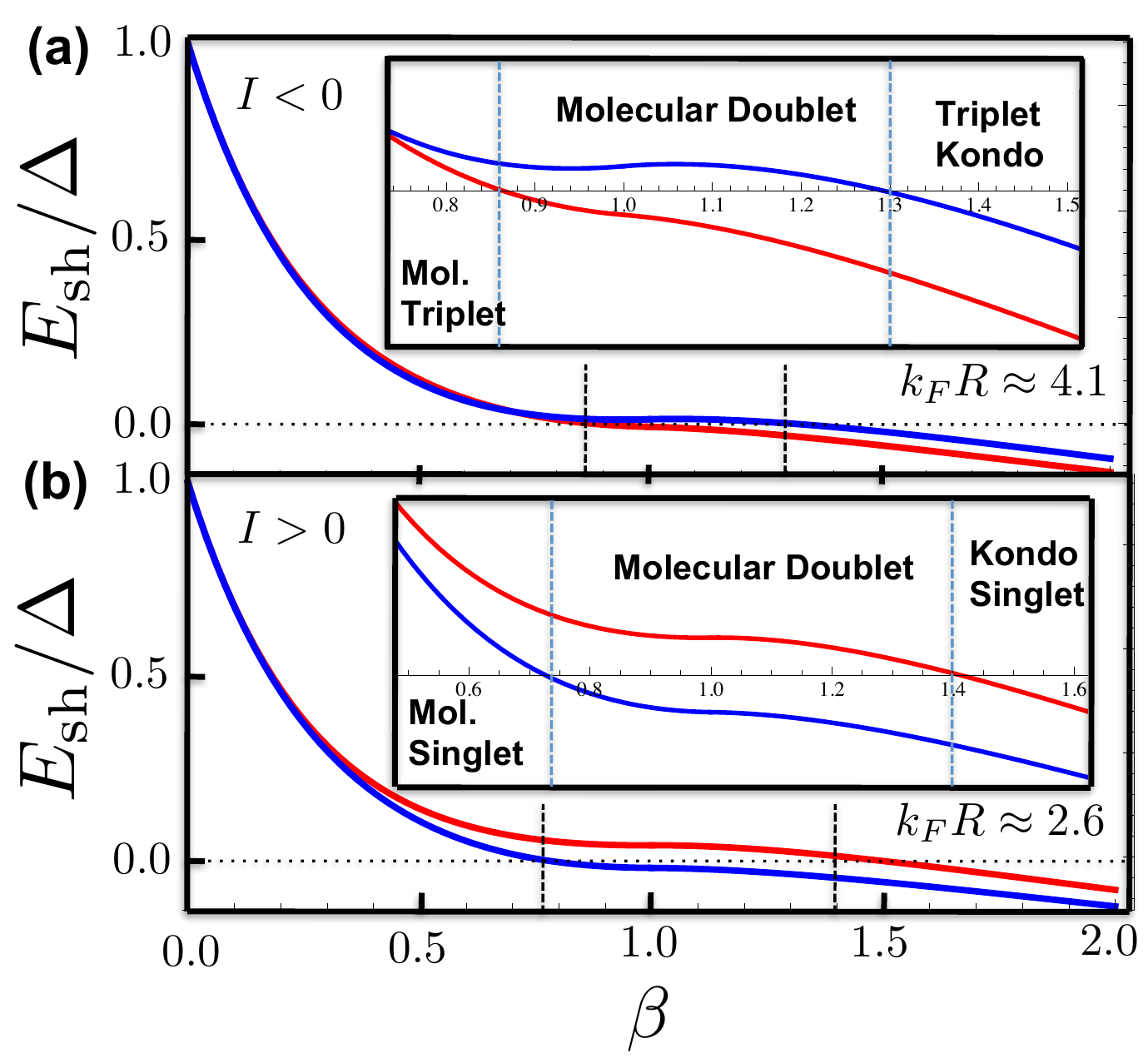}
\caption{Semi-classical molecular doublet phase transitions. (a) For $k_FR \approx 4.1$, the RKKY exchange is negative and the bound state energies are shown as one increases 
$\beta=JNS\pi/2$. 
At $\beta \approx 0.86$, the first bound state crosses zero and  a charge-parity transition from the molecular triplet phase to the doublet phase occurs. At $\beta \approx 1.3$, the second bound state crossing leads to the 
triplet Kondo 
phase.  (b) Analogous semi-classical results for $k_FR \approx 2.6$ where the exchange is positive. }
\label{fig:semiclassic}
\end{figure}

The existence of this novel molecular doublet phase can also be probed and confirmed in a semi-classical calculation where one extends the original Yu-Shiba-Rusinov calculation to the case of two classical magnetic impurities. 
Each magnetic impurity binds a Shiba state with wavefunction $\phi_{\text{sh}}(\text{\bf r}) \sim \frac{1}{\text{\bf r}} e^{-\text{\bf r}/\zeta |\sin(2\delta)|}$ and energy $E_{\text{sh}} = \Delta \frac{1-\beta^2}{1+\beta^2}$, where $\zeta$ is the coherence length, $\beta \equiv \tan(\delta) = JSN_0 \pi/2$ and $N_0$ is the density of states at the Fermi energy. 
Utilizing a two-impurity Green's function calculation \cite{Norm, Leonid}, we compute the energies of the hybridized Shiba bound states as poles of the $T$-matrix \cite{SuppInfo}. Picking two values of $k_FR$ (corresponding to ferromagnetic and anti-ferromagnetic exchange) we plot the bound-state energies as a function of $\beta$ (Fig.~\ref{fig:semiclassic}). In each case,  hybridization causes a single bound state to first cross $E_{\rm sh}=0$ leading to the formation of the molecular doublet phase. The second bound-state crossing then yields the transition to either the triplet Kondo phase ($I < 0$) or the Kondo singlet phase ($I>0$).

\begin{figure}[t] 
\centering
\includegraphics[width=0.9\columnwidth]{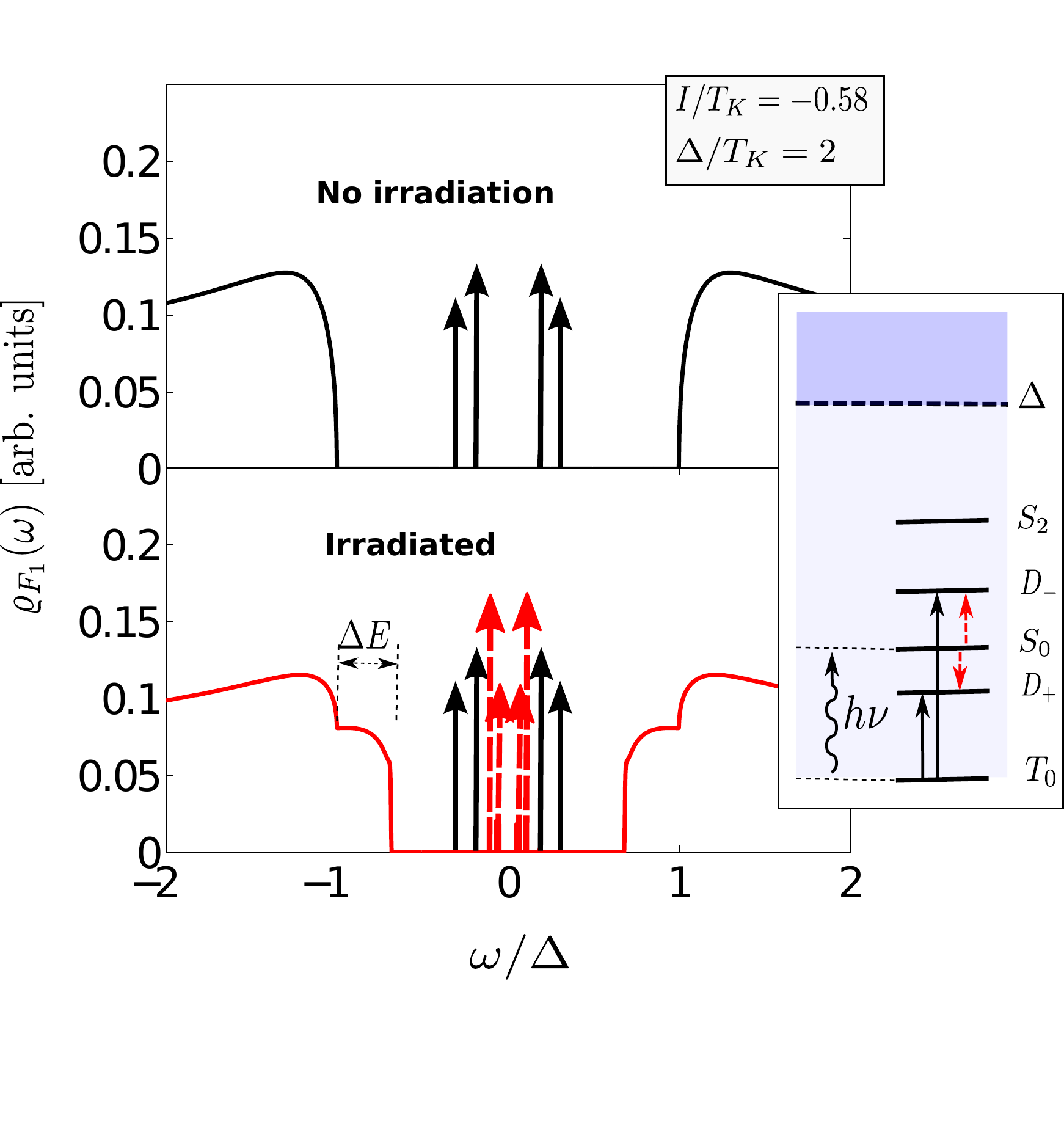}
\caption{
STM spectrum of one atom  of the Shiba molecule in the molecular triplet ($T_0$) phase. 
The $D_+$ and the $D_-$ states can both be observed as subgap Shiba transition lines (see top). 
Upon irradiation with a frequency matching the $T_0\to S_0$ transition $h\nu = \Delta E$ (right panel),
two additional subgap lines appear, and the gap shifts to lower values (bottom).  
 }
\label{fig:Shibaspectrum}
\end{figure}

\paragraph{Tunneling RF Spectroscopy} ---
The most direct observation of the various molecular Shiba states can be achieved by combining  RF spectroscopy  
with transport measurements. 
To this end, we determine the tunneling spectrum of the 
Shiba molecule by computing the spectral density of the so-called composite fermion, 
$F_1\equiv \bS_1 \cdot {\boldsymbol \s} \psi_1$. 
In the molecular triplet phase
($T_0$) 
both  $D_+$ and $D_-$ are visible in the tunneling spectrum and, correspondingly,  a double mid-gap STM resonance 
is predicted (see Fig.~\ref{fig:Shibaspectrum}). 
The dominant obstacle to observing such a resonance arises from thermal broadening; indeed, measurements of Mn and Gd impurities \footnote{Both Mn and Gd are high spin magnetic impurities. Adding in such effects (e.g. of single-ion anisotropy) is an interesting direction \cite{Zitko:2011p609}}
 on a single-crystal lead superconductor at $\sim 4$K are unable to resolve individual Shiba resonances \cite{Yazdani97}. However, operating at slightly lower temperatures ($\sim 500$mK) should reduce the linewidth to $\approx 0.14$meV, significantly smaller than the superconducting gap, $\Delta_{\rm Pb} = 1.55$meV. Such estimates are consistent with recent results which utilize a superconducting Niobium tip to explicitly resolve  multiple Shiba scattering channels \cite{Ji08, Ji:2010p607}. 
Much lower temperatures in the range of $T\sim 20\,{\rm mK}$ can be attained in mesoscopic circuits, where multiple Shiba states have indeed been 
resolved recently~\cite{Strunk}. 

Applying an additional RF field with a frequency matched to the $T_0\to S_0$ transition 
($\Delta E = h\nu$) 
allows one to  populate the  $S_0$ state \footnote{Parity must be broken to induce a  $S_0\to T_0$ transition, e.g. by an inhomogeneous magnetic field}. In this case, the
$S_0\to D_\pm$ transitions also become  active and visible (Fig.~5), while the 
tunneling gap   shifts from $\Delta\to \Delta- \Delta E$. 
In this way, one can detect the excited state  $S_0$ and its energy by  
investigating the RF-radiation-induced transport signal.

The transitions between the various phases and the 
corresponding STM spectra 
should also be observable  in double-dot spin-splitter devices. In particular, the tunneling ${\rm d}I/{\rm d}V$Ê
spectra  can be accessed by observing the transport   with  normal electrodes attached.
Similar to the case of a simple magnetic impurity, by approaching the phase boundaries between
$(D_+, S_2)$ or $(D_+, T_0)$, a single midgap excitation should get `soft' and cross zero. Interestingly,
the strength of the corresponding 
tunneling 
resonance displays a \emph{universal jump} at these transitions, 
$2\to 1$ and $3\to2$, respectively; this robust jump owes to a change in ground state degeneracy \cite{Tarucha}.

As a possible application, one can consider using the singlet states $S_2$ and $S_0$ as a quantum bit. These states are protected by the superconducting gap and, being singlets, they are insensitive to magnetic noise (including the hyperfine field of nearby nuclear moments) \cite{Yao14}. To have a direct transition between these states, both parity and particle-hole symmetry must  be broken sufficiently strongly; this can be achieved by  placing a single potential scatterer near one of the magnetic impurities, as may be possible in STM-type experiments \cite{ Yazdani97,Ji08, Ji:2010p607}.


We thank Leonid Glazman for insightful comments and lively discussions. This work is supported in part 
by the Hungarian research fund OTKA under grant Nos.
K105149, CNK80991,  the UEFISCDI grant
DYMESYS (ANR 2011-IS04-001-01, Contract No. PN-II-ID-JRP-2011-1),
  the `Iuventus Plus' project No. IP2011 059471,  the EU grant No. CIG-303 689, the DOE (FG02-97ER25308), the Harvard-MIT CUA, the ARO-MURI on Atomtronics, and the ARO MURI Quism program.
 Computing time at Pozna\'n Superconducting and Networking Center is acknowledged.

	
\bibliography{references}

\end{document}


\setcounter{figure}{0}
\renewcommand{\thefigure}{\Alph{figure}}

\begin{center}
{\bf Supplementary material for the paper
\\
``Phase diagram and excitations of a Shiba molecule''}
\\

by N. Y. Yao, C. P. Moca, I. Weymann, J. D. Sau, \\M. D. Lukin, E. A. Demler, and G. Zar\'and
\end{center}
\vskip 1cm

\noindent Here we provide details regarding the numerical renormalization group (NRG)  procedure, the semiclassical calculation, and the symmetries of the Hamiltonian. 

\section{Hamiltonian}

Let us start by specifying the explicit form of our Hamiltonian. We consider two spin 1/2 magnetic impurities, $\bS_{1}$  and $\bS_{2}$ embedded at positions $\bR_1$ and $\bR_2$ in an s-wave  superconductor. The Hamiltonian is given by 
\begin{eqnarray}
H_{\rm BCS} &= & \int \frac{d\bk}{(2\pi)^3}  \Big ( \sum_{\s} \xi_{\bk}c^{\dagger}_{\bk \s}c_{\bk \s}
+\bigl( \Delta c^{\dagger}_{\bk \uparrow}c_{-\bk \downarrow}^{\dagger} +h.c.
\bigr )\Big),
\nonumber
\\
H_{\rm int} &= &{J\over 2}\, \bS_1 \psi^{\dagger}_1{\boldsymbol \s}\,
\psi_1
+
{J\over 2}\, \bS_2 \psi^{\dagger}_2{\boldsymbol \s}\,
\psi_2\;,
\label{eq:H_int}
\end{eqnarray} 
where $\Delta = |\Delta| e^{-i \phi}$ stands for  the superconducting gap, and the $c^{\dagger}_{\bk\s}$ denote creation operators for  conduction electrons with spin $\s$ and momentum $\bk$. 
The latter  satisfy the usual anticommutation relations
$\{c_{\bk\s}, c^{\dagger}_{\bk'\s'} \}=(2\pi)^3\delta({\bk-\bk'})\,\delta_{\s\s'}$. The interaction part of the  Hamiltonian
is expressed in terms of   fields  $\psi^{\dagger}_1$ and $\psi^{\dagger}_2$,  which create conduction electrons at  positions $\bR_1$ and $\bR_2$, 
\begin{equation}
\psi_{i\s} \equiv  
\int \frac{d\bk}{(2\pi)^3} e^{i\bk \bR_i}\,c_{\bk \s}.
\end{equation}
These fields are normalized such that in the normal metallic phase ($\Delta = 0$), their local propagators decay asymptotically as 
$\langle\psi_{1\sigma}(t)\psi^\dagger_{1\sigma'}(0)\rangle \approx \varrho(0) \delta_{\s\s'} /it $, with $\varrho(0)$ being the local density of states per unit volume at the Fermi energy.
The Kondo temperature $T_K$ 
is then expressed in terms of the corresponding dimensionless coupling~\cite{Hewson}, $j=J\varrho(0)$ as 
$$
T_K \simeq D \,e^{-1/j}, 
$$
where $D$ is a high energy cut-off of the order of the conduction electrons' bandwidth.

\section{Mapping to the Wilson chain}

\subsection{Local action}

Since the coupling is restricted to the two sites $\bR_{1,2}$, both the ground state and the spin dynamics are completely determined by the normal and anomalous propagators of the fields $\psi_{j\s}$. 
As the system exhibits a real space mid-plane symmetry, it is justified to construct even/odd parity states. For that,
let us first introduce the normalized operators: 
\begin{equation}
a_{i\s}(\xi) \equiv  \frac 1 {\sqrt{\varrho(\xi)}} 
\int \frac{d\bk}{(2\pi)^3} e^{i\bk \bR_i}\,c_{\bk \s} \;\delta(\xi - \xi_\bk),
\end{equation}
satisfying the anticommutation relations 
$\{ a^\dagger_{i\s}(\xi), a_{i\s'}(\xi') \} = \delta_{\s\s'}\delta(\xi-\xi')$, and 
$
\{ a^\dagger_{1\s}(\xi), a_{2\s'}(\xi') \} = {\cal S}(\xi)\;  \delta_{\s\s'}\,\delta(\xi-\xi')
$. We can express the fields  
$ \psi_{i\sigma} = \int {\rm d}\xi\; \sqrt{\varrho(\xi)} \;a_{i\sigma}(\xi)\; $ in terms of these normalized operators. 
 The overlap factor ${\cal S}(\xi)$ measures the propagation amplitude of 
electrons of energy $\xi$ from site 1 to 2, and can be evaluated in 3D as
$
{\cal S}(\xi) = \frac{\sin(k R)}{k R}
\;,
$
with $\xi = k^2/2m - \epsilon_F$ and $R = |\bR_1-\bR_2|$. 
Parity allows us to define a new set of (properly orthonormal) even and odd fields:
$c_{e/o}(\xi)  \equiv (a_1(\xi) \pm a_2(\xi))/\sqrt{2N_{e/o}(\xi)}$, with 
$ N_{e/o} \equiv  1\pm {\cal S}(\xi)$,  which satisfy
 $\{ c_{a \alpha}^\dagger (\xi), c_{b\beta}(\xi') \} = \delta_{ab} \,\delta_{\alpha\beta}\, \delta(\xi-\xi')$, $\mbox{($a,b\in\{e,o\}$)}$.
It is these fields that we now use to build up $\psi_{j\s}$.  Their dynamics, as described by the equations of motion, 
are completely specified by the following action,
\begin{widetext}
\begin{eqnarray}
{\cal S}_0 &=& \sum_{a=e,o} \int dt \int d\xi \{
\bar {c}_{a\s}(\xi,t) (-i\partial_t +\xi)c_{a\s}(\xi,t)  +   [ 
\Delta\, \bar {c}_{a\uparrow}(\xi,t)  \bar {c}_{a\downarrow}(\xi,t)
+h.c.  ] \}.
\label{S_0}
\end{eqnarray}
One can easily check that this action indeed reproduces all normal and anomalous Green's functions of the 
fields $\psi_{i\sigma}$ and  $\psi_{i\sigma}^\dagger$.   The interaction term can then be written as 
\begin{eqnarray}
{\cal S}_{\rm int} & = & \frac 1 4
\int dt \int d\xi\,d\xi'  \Bigg\{ (\bS_1 +\bS_2) \Bigl(
j_{ee}(\xi,\xi') \, \bar {c}_{e}(\xi,t) {\boldsymbol \s} {c}_{e}(\xi',t) + "e \leftrightarrow o"\Bigr)
\nonumber
\\
&+& (\bS_1-\bS_2)  \Bigl ( j_{eo}(\xi,\xi') \, \bar {c}_{e\s}(\xi,t) {\boldsymbol \s} {c}_{o\s}(\xi',t) +
"e \leftrightarrow o"\Bigr)\Bigg \}, 
\nonumber
\end{eqnarray}
\end{widetext} 
with the exchange couplings defined as 
$j_{ab}(\xi,\xi') = J \sqrt{\varrho_a(\xi)\varrho_b(\xi') }$ and
$\varrho_{e/o}(\xi) = \varrho(\xi) (1 \pm {\cal S}(\xi))$.
Thus far our mapping is exact: one can use the actions above and reproduce, e.g., the three-dimensional RKKY 
interaction with little effort.

\subsection{NRG calculations}

In the numerical renormalization group approach, one typically replaces  the density of states $\varrho(\xi)$ by a flat density of states, 
$\varrho(\xi)\to n/2D$, with $n$ the density of unit cells, and also neglects the energy dependence of the couplings,  replacing them by their value at the Fermi energy, $
j_{e/o} = {1\over 2}\,j\, \left (1\pm {\cal S}^2\right )$, 
$j_{eo}  = j_{oe} = j_m = \sqrt{j_e\, j_o}$, 
with ${\cal S}= \sin(k_F R)/k_F R$ the overlap factor at the Fermi energy.
This procedure however, is   unjustified for quantities like the RKKY interaction, which is mediated by high energy electron-hole excitations, and depends on the details of the band structure, the dimensionality,  and the electronic dispersion. 
In fact, within this approximation, one would obtain an RKKY interaction
\begin{equation}
H_{\rm RKKY} \to  I_{{\rm NRG}} \bS_1\,\bS_2 = -2\ln{2}\,D\, j^2\, {\cal S}^2 \bS_1\,\bS_2\,,
\end{equation} 
that is always ferromagnetic, does not oscillate, and decays with the wrong exponent. 
To compensate this mistake, one must explicitly incorporate the RKKY coupling mediated by high 
energy carriers, and add a term $\Delta I  \, \bS_1\,\bS_2$ to the Hamiltonian~\cite{JonesVarma}. Then one can proceed in the 
usual way~\cite{Wilson.75} and map this zero-dimensional problem to the Wilson chain. Defining the local (dimensionless and normalized) operators
$$
f_{0,a\s} \equiv \frac {1} {\sqrt{2D}} \int \dxi\, c_{a\s}(\xi) 
$$
one obtains the following effective Hamiltonian (normalized by $D$): 
\begin{widetext}
\begin{eqnarray}
{\cal H}_0 = H_{\rm int}/D = (\Delta I/D)\, \bS_1\,\bS_2 + 
\sum_{a=e,o} j_a \;  \mathbf{s}_{a a}  \left( \mathbf{S}_1+\mathbf{S}_2 \right) 
+ j_m  \left( \mathbf{s}_{eo}  + \mathbf{s}_{oe} \right) \left( \mathbf{S}_1- \mathbf{S}_2 \right) \nonumber
\end{eqnarray}
\end{widetext}
with the spin operators defined as
\begin{eqnarray}
  \mathbf{s}_{aa'} = \frac{1}{2} \sum_{\s\s'} f_{0,a \s}^\dag \vec{\s}_{\s\s'} f_{ 0,a' \s'}.
\end{eqnarray}
The conduction electrons are represented by  semi-infinite Wilson chains
\begin{eqnarray}
H_{\rm band} &=& \sum_{a=e,o} \sum_{\s} \sum_{n=0}^\infty 
\Big \{
t_n \left(f_{ n,a\s}^\dag f_{n+1,a\s} +{\rm h.c.} \right) +  \left( \Delta  f_{n,a \uparrow}^\dag f_{n,a\downarrow}^\dag +{\rm h.c.} \right)
\Big \} \;.
\end{eqnarray}
This Hamiltonian displays   an obvious $SU_{\rm spin}(2)$ symmetry. In addition, however, 
it possesses an additional, hidden $U_c(1)$ symmetry,  
generated by the operator
\begin{equation}
\tilde Q_x = \frac 1 2\sum_{n, a}\big((-1)^n f_{n,a\uparrow}^{\dagger}f_{n,a\downarrow}^{\dagger} e^{i \phi}+h.c.\big),
\end{equation}
with $\phi$ denoting the phase of the superconducting order parameter. 
As we discuss in the next  section, this symmetry is related to the electron-hole symmetry of the Hamiltonian. 
We exploit this hidden $U_c(1)$  symmetry in the numerical calculations after performing  Bogoliubov and particle-hole 
transformations.


\subsection{Universal Shiba weight jumps}

Each  phase boundary in Fig.~2 corresponds to a level crossing of two sub-gap states. The boundaries of 
the doublet phase, $D_+$, in particular, correspond to crossings of the $D_+$ level with the states 
$S_2$, $S_0$, and $T_0$. All such transitions are `visible' in the STM spectrum in the sense that at the transition 
point, a Shiba peak moves to zero energy and crosses zero. Curiously, however, the \emph{amplitude} of the Shiba 
peak displays a universal jump at the transition, just as in the case of an ordinary Shiba transition~\cite{Bauer,Tarucha}.

\begin{figure}[t] 
\centering
\includegraphics[width=0.5\columnwidth]{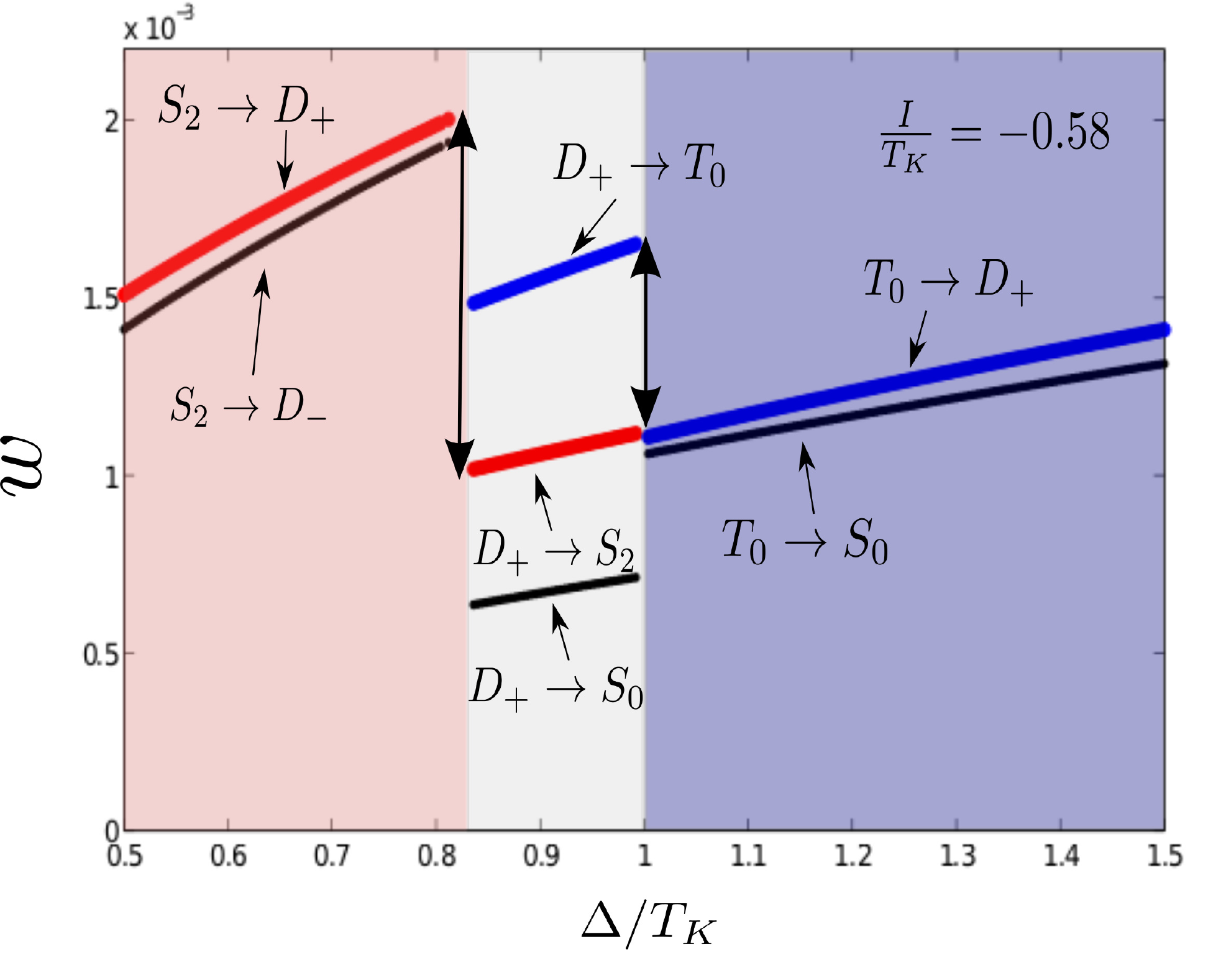}
\caption{
\label{fig:weights}
Weights of the Shiba transitions along the dashed line in Fig.~2 of the main text. The two-headed arrows indicate the 
universal 1/2 and 2/3 jumps for the states (central peaks)  crossing zero energy.}
\label{fig:phase_S_0}
\end{figure} 
The size of the jump is simply related to the degeneracy of the ground states on the two sides of the 
transition. At the phase boundary, $S_2\to D_+$, e.g.,  the even spin $S=1/2$ excited state gets `soft'
and crosses zero. The size of the jump follows from the assumption (which can be verified numerically) that 
the matrix elements $\langle S_2| F_{1\sigma}^\dagger | D_{+,\sigma'}\rangle $ and 
$\langle S_2| F_{1\sigma} | D_{+,\sigma'}\rangle $ are continuous at the phase boundary. In this case, 
on the singlet side of the transition, a spin $\uparrow$ electron  can always enter the superconductor
at the position of atom `1'   through the  
Shiba state with a probability amplitude, $\sim |\langle   D_{+,\downarrow}| F_{1\uparrow}^\dagger | S_2\rangle|^2 $.  On the doublet side, this transition turns into a hole-like process, where a hole enters the superconductor with transition amplitude
$\sim |\langle S_2| F_{1\uparrow} |   D_{+,\downarrow} \rangle|^2 $. In this case, however, the hole can only enter, if the 
supercondutor is in a state $\downarrow$, which happens with a probability $1/2$, since the two degenerate 
spin states of the doublet are equally probable.  As a result, the size of the Shiba peak
crossing zero is \emph{reduced} on the doublet side ($D_+$)  by a factor 1/2. 
Similarly, crossing the $D_+\to T_0$ boundary, the size of the central peak is prediceted to be suppressed 
by a factor $2/3$, corresponding to the doublet and triplet ground state degeneracies. 
The evolution of the weights of the Shiba states along the dashed line in Fig.~2 and the corresponding universal jumps
 are displayed in Fig.~\ref{fig:weights}.

\section{Hidden $SU_c(2)$ and $U_c(1)$ symmetries}

In this section, we demonstrate  on the particular example of a half-filled band of a simple cubic superconductor
how electron-hole symmetry implies the existence of a charge $U_c(1)$ symmetry. 
Let us  consider 
$H_{\rm BCS} =  H_0 + H_\Delta$, with the normal and superconducting parts defined as
\bea 
H_0 &=&  \int_{B.Z.} \frac{d\bk}{(2\pi)^3}  \sum_{\s} \xi_{\bk}c^{\dagger}_{\bk \s}c_{\bk \s}\;,
\\
H_\Delta & = &  \int \frac{d\bk}{(2\pi)^3}  \,\bigl( |\Delta| e^{- i\phi} c^{\dagger}_{\bk \uparrow}c_{-\bk \downarrow}^{\dagger} +h.c.
\bigr ).
\eea
Here $|\Delta|$ and  $\phi$  denote the amplitude and phase of the superconducting order parameter, respectively.
We set the lattice constant to unity, and assume a half-filled cubic lattice with nearest-neighbor hopping, and  corresponding dispersion, $\xi_{\bk} = -2t \sum_{i=x,y,z} \cos(k_i)$.

Let us now consider a mapping 
\be
\bk \to g(\bk) \equiv ({\mathbf{\pi}} -\bk)\mod 2\pi\ \label{eq:g1}
\ee
of the Brillouin zone to itself, and  the operators
\begin{eqnarray}
Q^z &=&  {1\over 2}\sum_{\s} \int \frac{d\bk}{(2\pi)^3} \Big ( c^\dagger_{\bk\s}c_{\bk\s}- {1\over 2}\Big),
\\
Q^+ &=&  \int \frac{d\bk}{(2\pi)^3}  c^\dagger_{\bk\uparrow}c^\dagger_{g(\bk)\downarrow }, 
\phantom{nnn}Q^- =  \big (Q^+\big )^\dagger.
\end{eqnarray}
For any $g(\bk)$ the  `charge' operators  $Q_x = (Q^+ +Q^-)/2$, 
$Q_y = (Q^+-Q^-)/2i$ and $Q_z$   can be shown to satisfy the standard $SU(2)$ charge algebra,
$\big[Q_i,\, Q_j\big] = i\, \epsilon^{ijk}\,Q_k$, while -- being spin 0 operators -- they commute with all components of 
the total spin  operator. Moreover,  since the dispersion satisfies
$$
\xi_\bk \equiv -\xi_{g(\bk)}\;
$$
by electron-hole symmetry, they can be shown to commute with  $H_0$ as well. Thus, $H_0$ possesses a hidden
charge $SU_c(2)$ symmetry, generated by  $\{Q_i\}$.

\begin{figure}[t] 
\centering
\includegraphics[width=0.4\columnwidth]{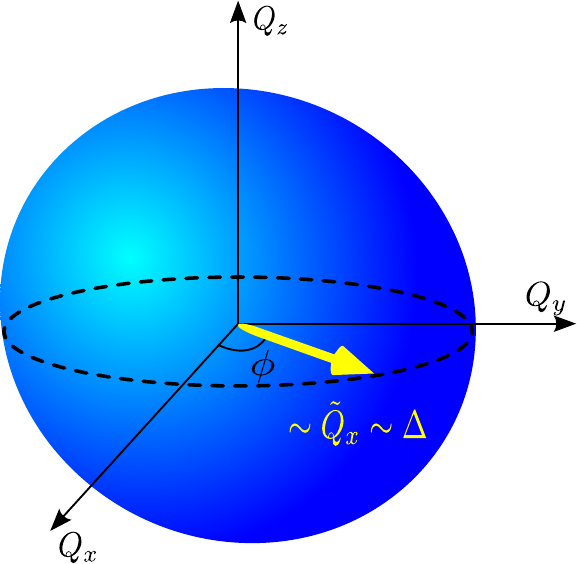}
\caption{
\label{fig:weights}
Bloch sphere representation showing the direction of the superconducting order parameter in ($Q_x$, $Q_y$, $Q_z$) coordinates. 
The phase  $\phi$ is the macroscopic phase of the condensate. $\tilde Q_x$ generates a rotation along the axis of 
the superconducting order parameter.}
\label{fig:Bloch}
\end{figure}

On the other hand, $H_\Delta$  certainly does not commute with 
the overall charge $Q= 2Q_z$, and therefore breaks  the expicit charge $SU_c(2)$ symmetry. 
Commuting it, however, with $\tQ^\pm \equiv e^{\pm i \phi} Q^\pm $ we obtain,  
\begin{eqnarray}
\big [ H_\Delta, \tQ^{+}\Big] &=& |\Delta| \sum_{\bk, \bq} \Big[ c_{-\bk\downarrow} c_{\bk\uparrow},\, c^{\dagger}_{\bq\uparrow} c^\dagger_{g(\bq)\downarrow}   \Big]\nonumber\\
& = & |\Delta| \sum_{\bk} \Big(  c^{\dagger}_{\bk\uparrow} c_{-g(\bk)\uparrow} +c^{\dagger}_{-g(\bk)\downarrow} c_{\bk\downarrow} \Big),
\nonumber\\
\big [ H_\Delta, \tQ^{-}\Big] &=& 
|\Delta| \sum_{\bk} \Big( c^{\dagger}_{-g(\bk)\uparrow} c_{\bk\uparrow}+ c^{\dagger}_{\bk\downarrow} c_{-g(\bk)\downarrow}    \Big)\, . 
\end{eqnarray}
Therefore,  the linear combination
\be
\tilde Q_x \equiv  (e^{i \phi} Q^+ +  e^{-i \phi} Q^-)/2  = \cos(\phi) Q_x + \sin(\phi) Q_y
\ee
commutes with $H_\Delta$, provided that   $g(\bk)$ fulfills 
\begin{equation}
g(\bk)= - g^{-1}(-\bk)\, , \label{eq:g}
\end{equation}  
a condition readily  satisfied by  Eqn~\eqref{eq:g1}. The operator $\tilde Q_x$  still commutes with the full BCS Hamiltonian 
$H_{\rm BCS}$, and generates a residual $U_c(1)$ symmetry. As illustrated in Fig.~\ref{fig:Bloch},
 $\tilde Q_x$ can be viewed as the generator of charge rotations along the superconducting order 
parameter.  

What remains to be shown is that  the interaction part of the Hamiltonian, $H_{\rm int}$ 
also commutes with $\tilde Q_x$. This naturally occurs since the local spin density operators 
${\boldsymbol \s}({\bR}_j) = \psi^\dagger_j {\boldsymbol \s}\psi_j$ 
 commute with all $Q_i$'s for the mapping defined in Eqn~\eqref{eq:g1} (for any lattice vector $\bR_j$). 
 
Thus, in a cubic, half-filled lattice, 
 our Hamiltonian exhibits a hidden electron-hole-type $U_c(1)$ symmetry -- in addition to the usual spin $SU_{\rm spin}(2)$  and  $e/o$ parity ($Z_2$) symmetries. Although, strictly speaking, this residual electron-hole symmetry 
 holds only under rather special conditions, it is expected to remain an \emph{approximate symmetry} as long 
 as the associated energy scale (typically of the order of the Fermi energy $E_F$) is much larger than the other scales of relevance, e.g.~the Kondo scale $T_K$, the superconducting gap $\Delta$, and the RKKY coupling, $I$. 

\section{Semiclassical Phase Diagram}

Here, we provide details of the semiclassical calculation of the molecular doublet phase. To probe the hybridization induced splitting of the Shiba bound states, we employ  a $T$-matrix formalism, $G_{\text{\bf k} ,\text{\bf k}'}(z) = G_{\text{\bf k}}^{(0)}(z) + G_{\text{\bf k}}^{(0)}(z) T_{\text{\bf k} ,\text{\bf k}'}(z)G_{\text{\bf k}'}^{(0)}(z)$  where $z= \epsilon + i 0^{+}$, $G_{\text{\bf k}}^{(0)}(z) = [z - (\xi_\text{\bf k}\tau^{z}+\Delta\tau^{x} ) ]^{-1}$ is the bare BCS Green's function ($\xi_\text{\bf k} = \frac{\text{\bf k}^2}{2m}$)
and  $T_{\text{\bf k} ,\text{\bf k}'}$ is the two-impurity $T$-matrix. To see the effects of Shiba wavefunction hybridization explicitly, we compute the bound state energies as a function of impurity separation. As usual,  these bound state energies $E_b$ can be extracted from  poles of  $\text{Tr}[G_{\text{\bf k} ,\text{\bf k}'}(z)] $.  More explicitly, $E_{\text{b}}$ is determined by 
\begin{align}
F(E_{\text{b}}) \equiv \text{Det}[1-SG(E_{b})] =0,
\end{align}
where  $S$ and $G$ are $4\times 4$ matrices given by,
$G_{ll'}(z)=\int \frac{{\rm d}^3\text{\bf k}}{(2\pi)^3} G^{(0)}_\text{\bf k}(z) e^{i  \text{\bf k} ( \text{\bf r}_l -  \text{\bf r}_{l'}) }$ and
$S_{ll'}=S_l\delta_{ll'} \otimes \tau^0$.
Here, $l$, $l'$ run over $\{L, R\}$, indexing the left/right impurity, $S_l$ characterizes the coupling strength and   $\tau^0$ represents the identity matrix in particle-hole space.
As one increases the effective exchange constant $\beta$, the bound state energies move toward zero. For large enough $\beta$, the hybridization induced splitting becomes appreciable and this pushes a single bound state to first cross zero energy. This parity changing transition is exactly that which leads to the  molecular doublet phase. 
To locate the transition, 
we  calculate the values of $\beta$ where each bound state crosses zero energy (for a particular $k_FR$), $F (E=0) =0$.
We begin by providing the analytic form for  $F(E=0)$,
\begin{align}
\hspace{-3mm}
F(E=0) &= \frac{\beta^4 e^{-\frac{2k_F R \Delta}{E_f}}}{(k_F R)^4} + (\beta^2-1)^2 + \frac{2 \beta^2 e^{-\frac{k_F R \Delta}{E_f}}(-1 + \beta^2 \cos (2k_F R))}{(k_F R)^2}.
\end{align}
Defining $\alpha \equiv  \frac{ e^{-\frac{k_F R \Delta}{E_f}}}{(k_F R)^2}$ we find a quartic equation in the effective exchange constant $\beta$, 
\begin{align}
\hspace{-3mm}
F(E=0) &= \alpha^2 \beta^4 + (\beta^2-1)^2 + 2\alpha \beta^2(-1 + \beta^2 \cos (2k_F R)) \nonumber\\
&= \beta^4 (\alpha^2 + 2\alpha \cos(2k_fR) +1) - \beta^2 (2\alpha +2) +1,
\end{align}
which yields  solutions of the form
\begin{align}
\hspace{-3mm}
\beta &= \left ( \frac{(2\alpha +2) \pm \sqrt{8\alpha - 8\alpha \cos(2k_F R)}}{2(\alpha^2 + 2\alpha\cos(2k_F R) +1)} \right)^{1/2}.
\end{align}
This provides an analytic semiclassical formula for the critical values of $\beta$ as a function of $k_F R$, where one of the 
 Shiba bound states crosses to negative energies. The first bound state crossing leads to the molecular doublet  transition while the second leads to either the triplet Kondo phase ($I<0$) or the Kondo singlet phase ($I>0$).

\bibliography{references}